# Performance Measurement of Security Academic Information System using Maturity Level

Endang Kurniawan, Amin Irmawan
Departement of Information Technology
Institut Teknologi dan Bisnis Muhammadiyah Wakatobi
Wakatobi, Indonesia

Imam Riadi
Departement of Information System
Universitas Ahmad Dahlan
Yogyakarta, Indonesia

Arusani
Departement of Entrepreneurship
Institut Teknologi dan Bisnis Muhammadiyah Wakatobi
Wakatobi, Indonesia

**Abstract:-** This study aimsto information security in academic information systems to provide recommendations for improvements in information security management by the expected maturity level based on ISO/IEC 27002:2013.

By using a qualitative descriptive approach, data collection and validation techniques with triangulation techniques are interviews, observation, and documentation. The data were analyzed by using gap analysis and to measure the maturity level determined 15 objective control and 45 security controls scattered in 5 clauses, the result of the research found that the performance of academic information system maturity level at level 2.

That is, the current level of maturity is below the expected maturity level, so it needs to be increased to the expected level.

## I. INTRODUCTION

Information is one of the most important organizational assets. With the rapid development of information technology, the possibility of disturbance of information equality is also increasing. Security issues are an important aspect of the use of information systems, often security is placed in the final order of the list that is considered important.

Currently, it is needed a good information system security system to keep the information owned by the company, although there is no perfect information system security because there is always a gap or a way to penetrate the security of an information system.

The presence of computer technology with its processing power has enabled the development of computer-based management information systems. By utilizing computer technology, obtained the benefits of ease of storing, organizing, and making retrieval of various data. Supported by the right software and hardware configuration, the company can build a reliable management information system and significantly influence the overall performance of the organization.

Generally, the occurrence that often occurs in college or academy is the limitations of data processing that starts from data processing for the examination screen of prospective students, the announcement of the results of prospective students who graduated, the re-registration process, both for prospective new students and students who have become students from college such height. The academic information system can be defined as a system designed to meet the needs of academics who want computerized education services to improve the performance, service quality, competitiveness, and quality of human resources it produces.

Above incident is one process which is an interaction between an internal part of college or academic represented by data processing or administration of data which have been arranged in such a way with certain process and procedure. It is expected with the existence of a data processing system between users who are students and processing that is part of the academic information systems that receive input from students and process it to conduct transactions activities lectures and administrative activities between students and college[1].

Investments issued for assessment and evaluation in the use of information technology should be considered. Based on some research that has been done, explained that the organization has begun to realize and start doing performance measurement and evaluation[2]. In the analysis of information technology, several frameworks refer to international information technology governance guidelines that have been widely accepted and proven to implement such as ISO 27001, COBIT, ITIL, and so on [3], which can be implemented by different organizational conditions.

The system can run according to its functions and needs, so an inspection process is needed to measure the performance of how the system is running. Security checks for information systems to function properly require the necessary standards.

Officially, there are no references that can be used as standards that can be chosen by an organization this can be selected according to the standard requirements for conducting information system security checks. The system can run according to its functions and needs then it is necessary to carry out a performance measurement process that is carried out through an inspection mechanism.

To be able to identify the extent to which an organization meets security standards, an identification framework can be used which is presented a maturity level that hat a grouping of organizational capabilities.

## II. MATERIALS AND METHODS

This chapter describes in detail the steps involved in being carried out systematically so that it can be used as a guideclear to solve the problem. The stages in this research can be seen in Figure 1.

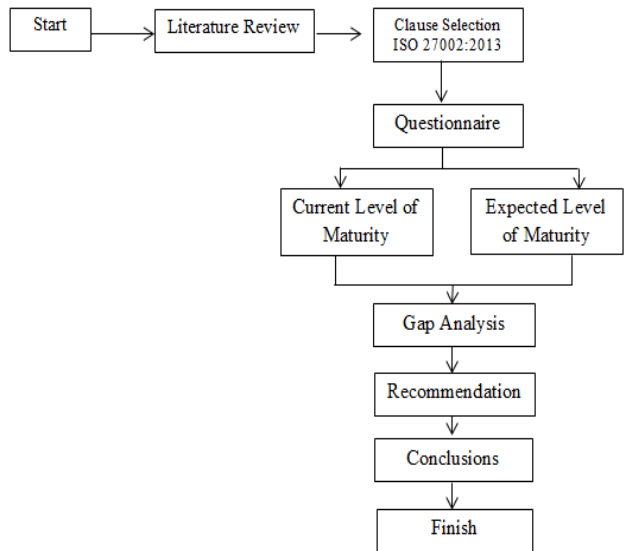

Fig. 1: Steps of Research Activities

In this research, a qualitative method is used, where the data obtained is based on the results of the questionnaire distributed to respondents.

The questionnaire is a list of questions given to respondents directly or indirectly. The questionnaire is an important aspect of research which consists of a series of questions to collect information from respondents.

Researchers usually use research questionnaires to collect data from respondents relatively quickly. A research questionnaire is the most effective tool to measure the behavior, attitudes, preferences, opinions, and intentions of research responses. Respondents only need to choose or answer a list of questions in the research questionnaire. Therefore, the research questionnaire can be considered a written interview that is determined based on the respondents' answers.

In the questionnaire in this study, the author made a list of questions according to the standards contained in ISO 27002 regarding instructions for implementing information security management which contains 5 criteria or clauses.

The scope of the audit is carried out by determining the purpose of the control to be used, this must be agreed upon in advance by the stakeholders in the form of a letter of agreement.

Within this scope, the organization must pay attention to the requirements for implementing and determining the risks if these controls are not met. Monitoring is designed to assure that management actions can ensure that business objectives are achieved and any risks that occur can be prevented and improved.

In Table 1, some clauses were agreed upon as inspection materials according to the ISO 27002 guidelines.

| Clauses | Description |
|---|---|
| 7 | Human Resources Security |
| 9 | Access Control |
| 11 | Physical and Environmental Security |
| 14 | System Acquisition, Development, and Maintenance |
| 16 | Information Security Incident Management |

Table 1: Clause of ISO 27002: 2013

The secondary data that the researcher used was obtained through literature studies such as books, journals, and proceedings.

The results of the questionnaires that have been distributed, then the data is processed using the maturity level method to determine the results of the calculation of the security level of the information system.

The scale used in this method uses the Guttmann scale. To get a definite answer, between conditions Yes - No, True-False, Positive-Negative, and others.

For this research, answers are provided Y (Yes) and N (No), where answer Y (Yes) is worth 1, and N (No) is worth 0, as shown in Table 2.

| Clauses | Description |
|---|---|
| 7 | Human Resources Security |
| 9 | Access Control |
| 11 | Physical and Environmental Security |
| 14 | System Acquisition, Development, and Maintenance |
| 16 | Information Security Incident Management |

Table 2: Example statement clause 11: Physical and Environmental Security

To process the data questionnaire, the researcher used the Microsoft Excel application.

All data from the questionnaire answers are entered into the table and then the maturity level of each process in each clause is calculated. In this study, the number of respondents who filled out the questionnaire was 10 respondents, as shown in Table 3.

The results of the analysis and interpretation of data processing and interviews can be used as research findings, and based on the results of these calculations can see the gap and determine what is the expected value to make a recommendation from each control objective carried out from each control objective that needs to be improved.

| No | Functional Structure | ∑ |
|---|---|---|
| 1 | Information Technology Division Manager | 1 |
| 2 | Assistant Manager | 1 |
| 3 | EDP Staff | 2 |
| 4 | Analyst Staff | 1 |
| 5 | Programmer Staff | 2 |
| 6 | Operator Staff | 3 |
| Total Respondents | | 10 |

Table 3: Respondents

## III. RESULTS

In the chapter, it is explained how the results of measuring the security performance of academic information systems according to the level of maturity are processed from the results of questionnaire data processing and interviews that have been carried out previously, based on the ISO 27002 standard.

### A. Summary of the Maturity Level

The calculation results from the distribution of the questionnaires are made on average from the respondents' answers according to the clause. Then the results of the maturity level can be obtained as follows:

a) Maturity level results clause 7: human resources security

Following the results of the calculation of the maturity level in process 7 of the ISO 27002 framework regarding human resources security, it is at the repeatable level but intuitive at position 2.39. This means the security system must be developed to a better stage.

From the data obtained, until now there has been no audit process security, but the policies issued by the organization have been applied to all existing departments.

The system has protected important documents from damage and loss, as shown in Table 4 and illustrated in graphical form.

The results of the calculation of the maturity level in clause 7 can be seen in Figure 2.

| Control Object | Description | Index |
|---|---|---|
| 7.1.1 | Rules and Responsibilities | 1.30 |
| 7.1.2 | Selection | 2.30 |
| 7.1.3 | Requirements and conditions to be met by employees | 2.75 |
| 7.2.1 | Management responsibilities | 1.65 |
| 7.2.2 | Education and information security training | 1.29 |
| 7.2.3 | Discipline process | 3.20 |
| 7.3.1 | Responsibility for dismissal | 2.76 |
| 7.3.2 | Asset return | 2.45 |
| 7.3.3 | Elimination of permissions | 3.80 |

Table 4: Calculation of Clause 7: Human Resources Security

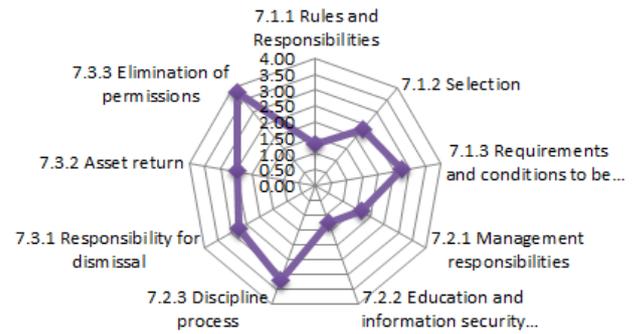

Fig. 2: Maturity Level of Clause 7: Human Resources Security

b) Maturity level results clause 9: Access Control

Following the results of the calculation of the maturity level in process 9 of the ISO 27002 framework regarding the access control, it is at Initial / Ad Hoc at position 1.49 which means that information security is currently not by process standards and must be improved.

The duties and responsibilities of information security must be carried out by all administrators of organizations that run academic information systems. Other parties outside the organization are not allowed to access information related to the organization, and can only access general information, as shown in Table 5 and illustrated in graphical form.

The results of the calculation of the maturity level of clause 9 of access control can be seen in Figure 3.

| Control Object | Description | Index |
|---|---|---|
| 9.1.1 | Access control policy | 2.15 |
| 9.2.1 | User registration | 2.10 |
| 9.2.2 | A privilege or special management | 1.25 |
| 9.2.3 | User password management | 1.20 |
| 9.2.4 | Review of user permissions | 1.30 |
| 9.3.1 | Use of passwords | 0.88 |
| 9.3.2 | Unattended user tools | 2.30 |
| 9.3.3 | Clear desk and clear screen policies | 2.14 |
| 9.4.1 | Network service usage policy | 1.24 |
| 9.4.2 | User authentication to connect out | 1.10 |
| 9.4.5 | Separation with the network | 0.50 |
| 9.4.6 | Control over network connections | 1.85 |
| 9.4.7 | Control of network routing | 1.40 |

Table 5: Calculation of Clause 9: Access Control

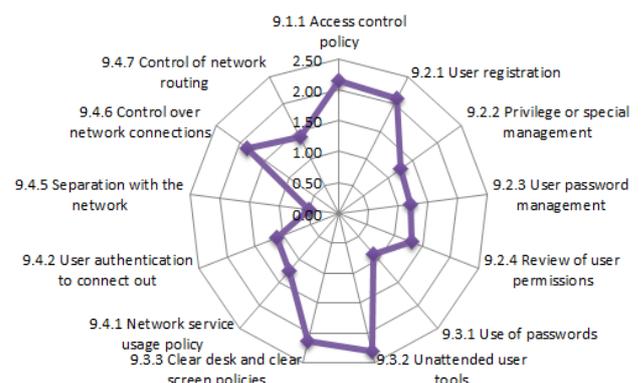

Fig. 3: Maturity Level of Clause 9: Access Control

c) Maturity Level Results Clause 11: Physical and Environmental Security

Following the results of the calculation of the maturity level in process 11 of the ISO 27002 framework regarding physical and environmental safety and is at the repeatable level but intuitive at position 2.41 which means this system must be developed to a better stage.

From the data obtained, until now there has been no audit process security, but the policies issued by the organization have been applied to all existing departments.

The system has protected important documents from damage and loss as shown in Table 6 and illustrated in graphical form.

The results of the calculation of the maturity level in clause 11 can be seen in Figure 4.

| Control Object | Description | Index |
|---|---|---|
| 11.1.1 | Physical security restrictions | 2.30 |
| 11.1.2 | Physical in control | 3.30 |
| 11.1.3 | Security office, space, and amenities | 3.00 |
| 11.1.4 | Protection against external attacks and environmental threats | 1.10 |
| 11.1.5 | Working in a safe area | 3.40 |
| 11.2.1 | Placement of equipment and protection | 3.30 |
| 11.2.2 | Supporting utilities | 2.34 |
| 11.2.3 | Security of wiring | 1.70 |
| 11.2.4 | Equipment maintenance | 1.34 |
| 11.2.5 | Safety equipment outside the workplace that is not hinted | 2.35 |

Table 6: Calculation of Clause 11: Physical and Environmental Security

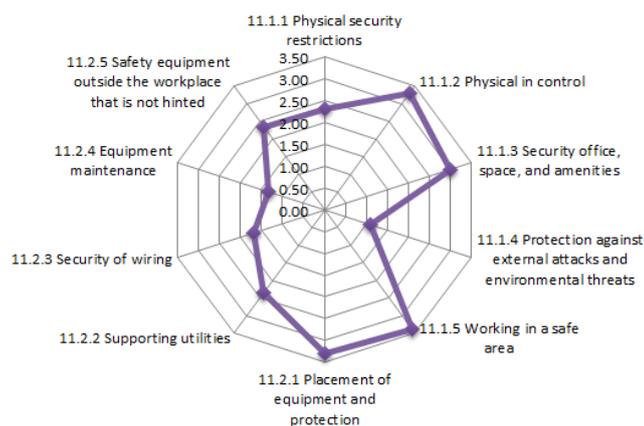

Fig. 4: Maturity Level of Clause 11: Physical and Environmental Security

d) Maturity Level Results Clause 14: Acquisition of Information Systems, Development, and Maintenance

Following the results of the calculation of the maturity level in process 14 of the ISO 27002 framework regarding Acquisition of Information Systems, Development, and Maintenance is managed and measurable at position 3.57 which information security is standard and must be documented and then published through training.

All information systems are designed and built by the Information Technology Department without any interference hands of outsiders as shown in Table 7 and illustrated in graphical form.

The results of the calculation of the maturity level in clause 11 can be seen in Figure 5.

| Control Object | Description | Index |
|---|---|---|
| 14.1.1 | Incorporate information security in the business continuity management process | 3.70 |
| 14.2.1 | Validate input data | 3.58 |
| 14.2.2 | Controls for internal processing | 3.55 |
| 14.2.4 | Validation of output data | 3.76 |
| 14.5.1 | Additional control procedures | 2.80 |
| 14.5.3 | Restrictions on software package changes | 3.80 |
| 14.5.4 | Weakness of information | 3.70 |
| 14.6.1 | Control of technical weakness (Vulnerability) | 3.65 |

Table 7: Calculation of Clause 14: Acquisition of Information Systems, Development, and Maintenance

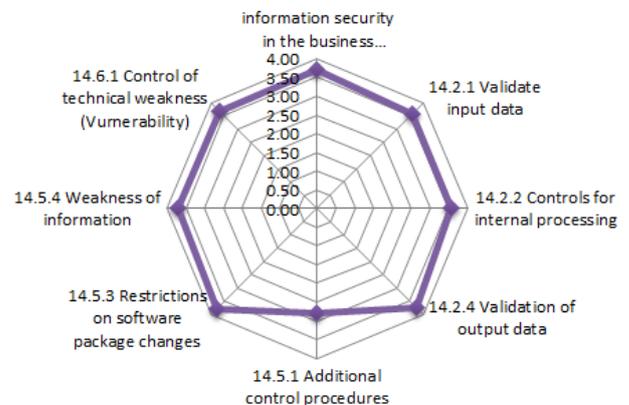

Fig. 5: Maturity Level of Clause 11: Acquisition of Information Systems, Development, and Maintenance

e) Maturity Level Results Clause 16: Information Security Incident Management.

Following the results of the calculation of the maturity level in process 16 of the ISO 27002 framework regarding Information Security Incident Management, it is at the repeatable level but intuitive at position 2.07. This means the security system must be developed to a better stage.

From the data obtained, until now there has been no audit process security, but the policies issued by the organization have been applied to all existing departments.

The system has protected important documents from damage and loss, as shown in Table 8 and illustrated in graphical form.

The results of the calculation of the maturity level in clause 16 can be seen in Figure 6.

| Control Object | Description | Index |
|---|---|---|
| 16.1.1 | Reporting information security events | 2.10 |
| 16.1.2 | Reporting security flaws | 2.25 |
| 16.2.1 | Responsibilities and procedures | 2.40 |
| 16.2.2 | Learn from information security incidents | 1.95 |
| 16.2.3 | Evidence collection | 1.65 |

Table 8: Calculation of Clause 16: Information Security Incident Management

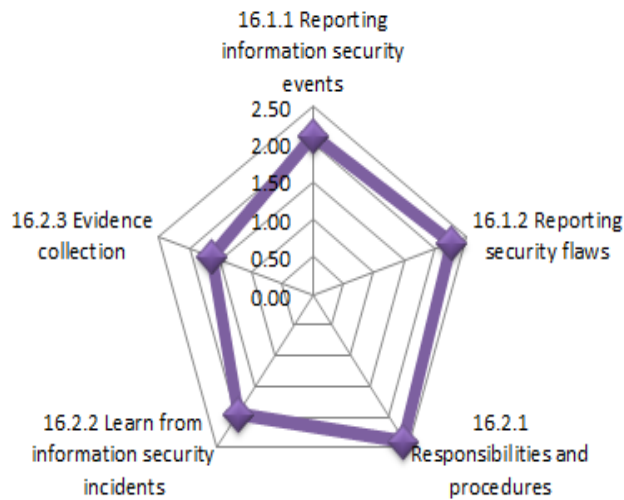

Fig.6: Maturity Level of Clause 16: Information Security Incident Management

From the results of the clauses, the maturity value obtained is calculated from the average respondents' answers to each clause contained in the ISO 27002 standard. Table 9 shows the results of the questionnaire calculation to obtain the maturity level of the academic information system.

The results of the calculation of the maturity level in the clause can be seen in Figure 7.

| Control Object | Description | Index | Level |
|---|---|---|---|
| 7 | Human Resources Security | 2.39 | 2 |
| 9 | Access Control | 1.49 | 1 |
| 11 | Physical and Environmental Security | 2.41 | 2 |
| 14 | System Acquisition, Development, and Maintenance | 3.57 | 3 |
| 16 | Information Security Incident Management | 2.07 | 2 |
| Average Maturity Level | | 2.39 | 2 |

Table 9. Result of Calculation Maturity

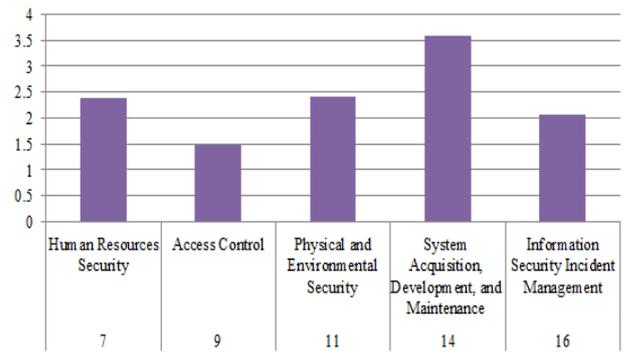

Fig. 7: Measurements graphs in maturity level

The results of the calculation to get the average value of information security control with maturity level in academic information systems is 2.39.

The maturity level, in this case, can be said that information security is at level two, which is repeatable but intuitive.

This means that the organization has a pattern that is repeatedly carried out in carrying out activities related to the management of information technology governance, but its existence is not well defined and there are still formal inconsistencies.

The results that can be seen in Table 9, for each process in the checked clause, can be seen in Figure 7 above.

B. *The Gap of Value Maturity Level*

The results of the current maturity calculation show that the security of the system is at level 2.39, this means that the system is in the repeatable but intuitive level where the expected level of security is at level 5 because the organization hopes that the readiness in the field of security policies, procedures, and processes for system security control information can run according to the needs of the organization. This can be seen in Table 10, a comparison of the current situation after the security level check with the expected situation.

| Control Object | Description | Maturity Cur. | Exp. | Gap |
|---|---|---|---|---|
| 7 | Human Resources Security | 2.39 | 5 | 2.61 |
| 9 | Access Control | 1.49 | 5 | 3.51 |
| 11 | Physical and Environmental Security | 2.41 | 5 | 2.59 |
| 14 | System Acquisition, Development, and Maintenance | 3.57 | 5 | 1.43 |
| 16 | Information Security Incident Management | 2.07 | 5 | 2.93 |
| Average Gap | | 2.39 | 5 | 2.61 |

Table 10: Result of Maturity Level Gap

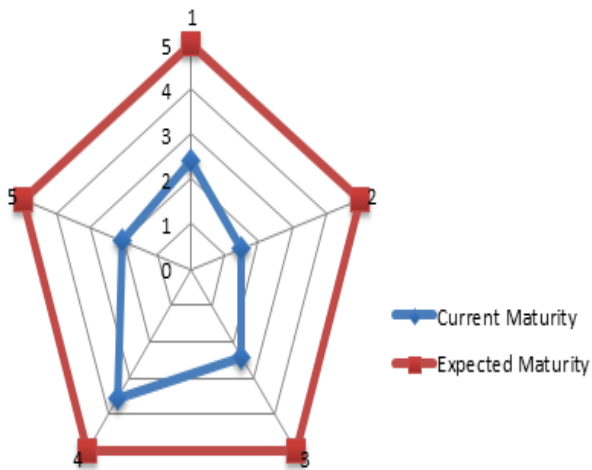

Fig. 8: Maturity Level Gap

Based on Table 10 above, can be seen the distance between the current condition and what is expected by the organization in each clause.

Where clause 7 has a gap of 2.61, clause 9 has a gap of 3.51, clause 11 has a gap of 2.59, clause 14 has a gap of 1.43, and clause 16 has a 2.93 gap.

After knowing the value of the gap in each clause, the next step of the resulting gap value is then summed. The daily sum of each of these gap values, they are averaged to get the overall gap value.

After calculating the overall gap value, the resulting value is 2.61 between the maturity of the current condition and the expected condition.

From the final results of the measurement of the security performance of the information system, it can be seen that the gap is very high, so it is necessary to adjust each control according to the ISO 27002 standard to focus more on weak controls.

The ratio of the value of the current maturity level and the value of the expected maturity level is depicted in Figure 8.

From Figure 8, itcan be seen that the current maturity level is represented by a blue line, while the expected maturity level is represented by a red line. From this line, it can be seen that the expected maturity of information system security at level 5 has been changed and continues to experience improvements in the information system security process against system changes that are carried out.

The selection of clauses that need to be changed is based on the results of the analysis where the control objective values are spread over a range of values from 1 to 3.

From this, it can be explained that the security level which has the lowest value from the gap analysis is 3.51 in clause 9 with the current information system security maturity level at 1.49.

The highest value is in clause 14 with a maturity value of 3.57 so it has the lowest analysis gap of 1.43. Thus, the higher the gap in the clause, the higher the risk to information system security. This explanation can be seen in Figure 8, from this line it can be seen that the maturity of information system security is expected at level 5 to have made changes and continues to experience improvements to the current system.

*C. Recommendations*

Based on the analysis performance measurement of the academic information system that has been done, the gap maturity level found in the control objectives in clauses 7, 9, 11, 14, and 16 can be found to be the solution by the organization regarding the ISO 27002 literature, especially on the Maturity level. The preparation of recommendations as a result of performance measurement of the application of academic information systems is emerging after comparison of what should be done with the ongoing process of the organization. From these results are then given recommendations that can be used to improve the process of information systems in the future. Here are the recommendations are given as an enhancement step:

- Based on the results of data analysis on ISO 27002 clause 7 Human Resources Security is recommended to remove the access rights when there are employees who are transferred to another division or when resigned. In addition, the organization must have official procedures for the administration of employees which includes registration, editing, and deletion of employee names.
- Based on the results of data analysis on ISO 27002 clauses 9 Access Control, the need to apply written rules on the obligation of all employees who have a user- id and password must change the password regularly for example 6 months. Regarding the password itself, it is advisable to have a combination of numbers and letters of no less than 8 characters meant that passwords have high quality.
- Based on the results of data analysis on ISO 27002 clause 11 Physical and Environmental Security need to install CCTV and Fingerprint door lock in the server room to prevent theft, destruction of a server, and access by unauthenticated peopled.
- Based on the results of data analysis on ISO 27002 clauses 14 Acquisition, Development, and Maintenance of Information Systems need a validation checking procedure that is entered directly into the academic information system to detect the damage of information through processing mistakes or deliberate actions.
- Based on the results of data analysis on ISO 27002 clauses 16, Information Security Incident Management needs to build a system that performs monitoring, warning, and vulnerabilities used to detect information security incidents so management can immediately resolve the incident quickly and effectively.

## IV. DISCUSSION

Based on the result of performance measurement of security academic information system using maturity level in this research determined 15 objective controls and 45 security control spread in 5 clauses ISO 27002 used.

The results obtained from the measurement of the level of maturity for information security is level 2 (repeatable but intuitive), results of the questionnaire management to obtain an average value for all of the clauses is 2,39 range of 0 to 5, and the value of the gap between current security conditions and the condition of the expected 2.61.

From this value can be concluded that the security information on the second level is repetitive but intuitive. Thus the results of the performance measurement academic information system have a pattern that is repeatedly performed in conducting activities related to the management of information technology governance, but its existence has not been well defined and that is still happening formal inconsistency.